\documentclass[a4paper,twocolumn,11pt,accepted=2024-01-15]{quantumarticle}
\pdfoutput=1
\usepackage[utf8]{inputenc}
\usepackage[english]{babel}
\usepackage[T1]{fontenc}
\usepackage{amsmath}
\usepackage{amssymb}
\usepackage{hyperref}
\usepackage{listings} 

\usepackage{tikz}
\usepackage{lipsum}
\usepackage{graphicx}
\usepackage{hyperref}
\usepackage{microtype}
\usepackage{bbm}
\usepackage{color}
\usepackage{subcaption}
\usepackage{cleveref}
\usepackage{graphicx}
\usepackage[export]{adjustbox}
\usepackage{float}
\usepackage{xcolor}
\usepackage{soul}

\newenvironment{code}{\samepage\vspace{-.3cm}}{\vspace{.4cm}}
\newcounter{codecount}
\newcommand\clabel[1]{\refstepcounter{codecount}\label{#1}}
\newcommand\ccaption[2]{\vspace{-.25cm}\small\noindent Code sample \ref{#1}: #2}
\newcommand{\coderef}[1]{code sample \ref{#1}}
\usepackage{amssymb,amsmath}
\usepackage{ifxetex,ifluatex}
\ifnum 0\ifxetex 1\fi\ifluatex 1\fi=0 
  \usepackage[T1]{fontenc}
\else 
  \ifxetex
    \usepackage{mathspec}
  \else
    \usepackage{fontspec}
  \fi
  \defaultfontfeatures{Ligatures=TeX,Scale=MatchLowercase}
\fi
\IfFileExists{upquote.sty}{\usepackage{upquote}}{}
\IfFileExists{microtype.sty}{%
\usepackage{microtype}
\UseMicrotypeSet[protrusion]{basicmath} 
}{}
\PassOptionsToPackage{hyphens}{url} 
\hypersetup{
            pdfborder={0 0 0},
            breaklinks=true}
\usepackage{color}
\usepackage{fancyvrb}

\DefineVerbatimEnvironment{Highlighting}{Verbatim}{commandchars=\\\{\}, fontsize=\footnotesize}
\usepackage{framed}
\usepackage{setspace}

\definecolor{shadecolor}{RGB}{235,235,235}
\newenvironment{Shaded}{\begin{snugshade}\setstretch{0.85}}{\end{snugshade}}
\newcommand{\KeywordTok}[1]{\textcolor[rgb]{0.13,0.29,0.53}{\textbf{#1}}}
\newcommand{\DataTypeTok}[1]{\textcolor[rgb]{0.13,0.29,0.53}{#1}}

\newcommand{\FloatTok}[1]{\textcolor[rgb]{0.53,0.00,0.00}{#1}}
\newcommand{\ConstantTok}[1]{\textcolor[rgb]{0.00,0.00,0.00}{#1}}

\newcommand{\StringTok}[1]{\textcolor[rgb]{0.31,0.60,0.02}{#1}}

\newcommand{\ImportTok}[1]{#1}
\newcommand{\CommentTok}[1]{\textcolor[rgb]{0.56,0.35,0.01}{\textit{#1}}}

\newcommand{\FunctionTok}[1]{\textcolor[rgb]{0.00,0.00,0.00}{#1}}

\newcommand{\ControlFlowTok}[1]{\textcolor[rgb]{0.13,0.29,0.53}{\textbf{#1}}}
\newcommand{\OperatorTok}[1]{\textcolor[rgb]{0.81,0.36,0.00}{\textbf{#1}}}
\newcommand{\BuiltInTok}[1]{#1}

\newcommand{\NormalTok}[1]{#1}
\ifx\paragraph\undefined\else
\let\oldparagraph\paragraph
\renewcommand{\paragraph}[1]{\oldparagraph{#1}\mbox{}}
\fi
\ifx\subparagraph\undefined\else
\let\oldsubparagraph\subparagraph
\renewcommand{\subparagraph}[1]{\oldsubparagraph{#1}\mbox{}}
\fi

\makeatletter
\def\fps@figure{htbp}
\makeatother

\usepackage[numbers]{natbib}

\newcommand{\vect}[1]{\mathbf{#1}}

\newcommand{\listTitle}[1]{\textbf{#1}}

\begin{document}

\title{BosonSampling.jl: A Julia package for quantum multi-photon interferometry}

\author{Benoit Seron}
\email{benoitseron@gmail.com}
\affiliation{Quantum Information and Communication, Ecole polytechnique de Bruxelles, CP 165/59, Universit\'e libre de Bruxelles (ULB), 1050 Brussels, Belgium}

\author{Antoine Restivo}
\email{antoine.restivo@ulb.be}
\affiliation{Quantum Information and Communication, Ecole polytechnique de Bruxelles, CP 165/59, Universit\'e libre de Bruxelles (ULB), 1050 Brussels, Belgium}

\maketitle

\begin{abstract}

We present a free open source package for high performance simulation and numerical investigation of boson samplers and, more generally, multi-photon interferometry.
Our package is written in Julia, allowing C-like performance with easy notations and fast, high-level coding. 
Underlying building blocks can easily be modified without complicated low-level language modifications.
We present a great variety of routines for tasks related to boson sampling, such as statistical tools, optimization methods and classical samplers. Special emphasis is put on validation of experiments, where we present novel algorithms. This package goes beyond the boson sampling paradigm, allowing for the investigation of new interferometric behaviours such as bosonic bunching.
\end{abstract}

\section{Introduction}

Quantum computing holds strong hopes for a profound change of computational paradigm and power in the coming years \cite{feynman2018simulating, shor1994algorithms, Tillmann_2013}. A large number of experimental and theoretical efforts took place in the last decades, leading to the fast growth in experiments' complexity. While a general purpose, universal quantum computer seems so far out of reach, non-universal models already claim a quantum advantage for specialized tasks on a quantum system that cannot be simulated in a reasonable amount of time with the most advanced classical super-computers \cite{preskill2018quantum, preskill2021quantum}. One of these restricted models is that of boson sampling. In their seminal paper \cite{aaronson2011computational}, Aaronson and Arkhipov describe a quantum version of Galton's board where $n$ bosons - generally photons - are sent through a $m$-modes linear interferometer implementing a unitary transformation $U$. 
The task consists in sampling output mode patterns of particles, such as finding $2$ photons in the first mode, none in the second, etc. 
Depending on the importance of the noise sources from the experimental components, the output distribution $\mathcal{D}$ can be hard or easy to sample from \footnote{One specific requirement for the proof of hardness is that the number of modes scales as $n^6$, while the hardness is strongly thought to hold even with $m \approx n^2$. We refer to Boson Sampling as the sampling task described above, whether it is in the quantum advantage regime or not.}. 
By hard we understand that it takes a super-polynomial number of steps to generate a sample from $\mathcal{D}$ on a classical computer. Indeed, for modest experimental noise, AA prove that this task remains hard, under widely believed conjectures in complexity theory. However, when sufficiently strong noise is present, e.g. due to partial distinguishability or particle loss, then classical algorithms can sample from $\mathcal{D}$ efficiently \cite{rahimi2016sufficient, renema2018efficient, oszmaniec2018classical, garcia2019simulating, brod2020classical,renema2018classical, shchesnovich2019noise}.


\begin{figure}[t]
    \centering
    \includegraphics[width=0.45
    \textwidth]{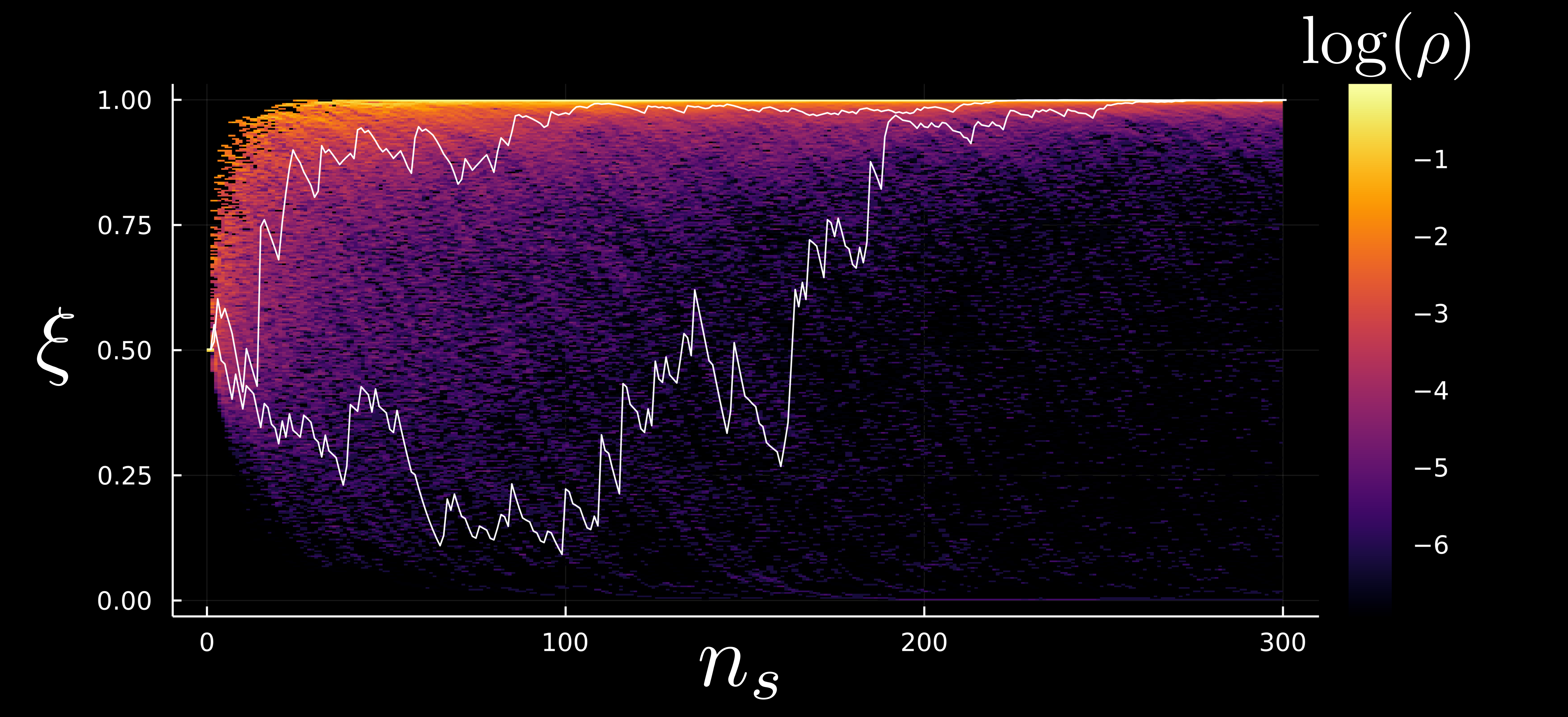}
    \caption{Validating boson sampling using Bayesian methods (see Sec. \ref{sec:validation}). Is displayed the confidence $\xi$ that the input is made of $10$ indistinguishable photons, averaged over $n_{trials} = 500$, as more samples $n_s$ are provided to the validation protocol. The color scale represents the density of $\xi$ curves. White curves are examples of validation runs.}
    \label{fig:validation}
\end{figure}


Boson sampling sparked a great interest both from theoreticians and experimentalists. Various alternative schemes were constructed, such as Scattershot and Gaussian boson sampling (GBS) \cite{Bentivegna_2015,Hamilton_2017} to alleviate the technical difficulties in generating single photons. These efforts culminated in multiple claims of quantum advantage in GBS \cite{zhong2020science, zhong2021phase, madsen2022xanadu, deng2023gaussian}, therefore questioning the validity of the extended Church-Turing thesis \cite{hangleiter2023computational}. Standard boson sampling saw experimental implementations with $n = 20$ photons in $m = 60$ modes \cite{wang2019boson}. Other experimental platforms than photonics were also considered \cite{robens2022boson, young2023atomic}.

These factors imply a great need for optimized, high performance numerical simulations of multi-photon interferometry. As the fight for quantum advantage is played between quantum devices and classical computers, scalable, versatile and fast implementations of the boson sampling algorithms become even more important. 
In particular, an adaptive framework for the boson sampling community is of great interest given the speed of growth of this field of research. Indeed, 
many new boson sampling paradigms were introduced theoretically, and a great variety  of experimental techniques were refined, enhancing the demand for evolutive code bases.

In this paper we expose our package, \textsc{BosonSampling.jl}, written in the Julia programming language to tackle classical tasks related to boson sampling. 
It intends to solve the challenges discussed above by providing an adaptable and extendable framework for multi-photon interferometry designed to study new boson sampling paradigms.
The choice of Julia solves the two-language problem: instead of using a high-level wrapper (e.g. Python) with time-critical functions written in a low-level languages (e.g. C, Fortran), Julia is fast out-of-the-box while being easy to write.
As a byproduct, all functions are fast, and not only time-critical ones.
This is especially relevant for time-crunched researchers who are not professional-programmers but still require a fast execution time for this type of computationally intensive research.
The package is evolutive and obeys the Open Source Software standards. It welcomes contributions from the community through its GitHub portal \cite{BosonSamplingGithub}. A complete documentation as well as a pedagogical tutorial are provided \cite{BSdoc}.




The paper is structured as follows. We first explain how the specific choice of the Julia programming language is, in the eyes of the authors, optimal for both theoreticians and experimentalists and how it affects the structure of our package and benefits users.
In the second section of this paper, we explore the capabilities of our codebase and expose how our package is structured, including more specific descriptions and examples in the third section. 
It contains worked-through examples of well-known physical phenomena such as the Hong-Ou-Mandel effect. We also show how one can make use of the package architecture to take into account new sources of noise for instance. 
Finally, we compare our package to other existing software and discuss possible future features.



\subsection{Julia}
\label{julia}
Julia is a high-level and dynamic programming language originally designed for technical computing, in particular, the simulation of physical systems. 
It is catching more and more attention from the scientific community and has been adopted in several high class projects such as working with dynamical systems \cite{Datseris2018}, satellites simulations \cite{satellite}, the Celeste project (1.54 petaFLOPS) \cite{regier2016learning}, the Climate Modelling Alliance \cite{OceananigansJOSS}, as well as NASA who saw a 15.000 times speed improvement over their previous Simulink/MATLAB code \cite{Anantharaman2020}. 
The field of quantum information and computation is already well covered, with several packages such as \textsc{Yao.jl} and \textsc{Quantum Optics.jl} rivalling in scope and efficiency with the leaders of the market \cite{Gawron2018,yao,quantumOptics,cumulants}.

The main strength of Julia, in particular for scientists, is its goal to solve the "two-languages problem".
It is very common that code is first prototyped in a high-level, easy to write, language such as Python and then made fast to execute in low-level, and hard to write languages such as C++.
Julia allows the coding to be convenient and efficient, while being as fast as lower level languages (such as C and Fortran) out of the box. 
This performance comes from its just-in-time (JIT) compilation and specific type architecture, more akin to functional programming than Object Oriented Programming (OOP) found in most languages (C++, Java, Python,...).
This bottom-up approach is especially suited for scientists, whose code evolve 
organically as results are found, without a set roadmap ("proofs then theorems") as would be preferred for OOP programming \cite{julia}.

Another strength of Julia's type structure is multiple dispatch: functions are written without strong restrictions to the type of their arguments, and can be reused with custom types with no modifications \cite{julia_high} (provided that the function makes sense for these new types). 
For instance, a function computing the square root would be restricted to specific types (say, \texttt{Float64}) in strongly typed languages. 
A new type, say \texttt{BigFloat}, would require a rewriting of the function, while in Julia the same square root algorithm will work automatically.
This design philosophy, is, once again, a blessing for scientists who may not know in advance what the future of their code looks like.

\section{Contents and architecture}

\subsection{Package overview}
\label{overview}

Our package, \textsc{BosonSampling.jl} together with its paired package \textsc{Permanents.jl}, provides tools to classically simulate multi-photon interferometry experiments. This includes the well-known cases of standard boson sampling, scattershot and Gaussian boson sampling.

Our platform is aimed at high-performance computations for scientists. This is reflected in the choice of the language, Julia, and the modular architecture. Our goal is to make it both easy and fast to implement new models while maintaining all the provided methods and tools adapted to the new features with fast execution time. 

The package is evolutive, and welcomes outside contributions from the community. It is intended to be a toolkit for both experimentalists and theoreticians, allowing to easily bridge between new theoretical models and their realistic implementation. 

We provide state-of-the-art tools addressing relevant boson sampling problems. Let us now give an overview of these tools made available.

\begin{enumerate}
    \item \listTitle{Boson-samplers}, which allow to get samples as if performing an experiment. These samplers are based on the best known algorithms from the literature. Those samplers, include experimental noise such as photons' partial distinguishability and loss. 
    This noise is critical as it will affect the classical hardness of simulating a given experiment. 
    
    \item \listTitle{Event probability} computation routines, allow to observe specific input/output patterns. Special interest is given to noise, and in particular to a fast implementation in the case of partial distinguishability.
    
    \item \listTitle{Bunching} tools, aimed at investigating the tendency of bosons to bunch and of fermions to anti-bunch, as exemplified by the Hong-Ou-Mandel effect. We provide functions to investigate the link between bunching and matrix permanent conjectures \cite{seronBosonBunching, shchesnovich2016universality}. 
    
    \item \listTitle{Validation} tools for boson-samplers. These allow, given experimental as well as black-box samples, to discuss the degree of multi-photon interference observed in a boson sampler device, and to estimate the level of noise. This gives indications as to whether the device outputs samples that are classically hard to obtain, i.e. if quantum advantage can be claimed. 
    Standard tools from the literature are unified in a recent validation protocol \cite{validation} along with the common method of correlators \cite{walschaers2016statistical, walschaers2018statistical}, suppression laws \cite{tichy2014stringent, dittel2018totally, viggianiello2018experimental, crespi2015suppression}, and full bunching \cite{shchesnovich2016universality, shchesnovich2021distinguishing} and also allowing to recover marginal probabilities \cite{renema2020marginal}. 
    
    
    \item \listTitle{Optical circuits} built from optical elements. While in the standard boson sampling problem, networks, described by a unitary matrix $U$, are chosen from the Haar measure, specific interferometers are also implemented such as the Fourier and Hadamard transformations. 
    In addition, a suite of linear-optical elements is provided, so that networks can be constructed as in an experimental setting.
    
    \item \listTitle{Optimization} routines to maximize cost functions over unitary matrices. Maximizing certain properties of interferometers, such as photonic bunching, can be convenient for the design of experiments. We include algorithms from Ref \cite{abrudan2009conjugate} that can optimize cost-functions on the Haar measure. 
    
    \item \listTitle{Time-bin interferometry} types and circuits to simulate the experimental boson sampling proposal of \cite{motes2014scalable}, where photons are separated in time bins and sent through a time-dependant beam-splitter and delay lines. 
    
   \item \listTitle{Applications of boson sampling} such as quantum cryptographic tools \cite{nikolopoulos2016decision, nikolopoulos2019cryptographic, wang2023} and cryptocurrency proof-of-work miners \cite{singh2023proofofwork}.

   \item \listTitle{Fundamental many-photon interference} investigation tools, such as generic (potentially entangled) input/output multiphoton interferometry \cite{shchesnovich2016universality}, generalized matrix function analysis \cite{Zhang+2016} and counter-example search \cite{shchesnovich2016permanent} and quantum photo-thermodynamics tools \cite{Becker_2021, anguita2022quantum}. Fermion sampling is also implemented.
\end{enumerate}

\subsection{Code architecture}

The two main procedures achievable with this package are the simulation of photonic experiments and validation for multi-photon quantum experiments. We will first describe the simulation architecture and then the validation procedure.

\subsubsection{Simulation}

\begin{figure}[h]
    \centering
    \includegraphics[width=0.5\textwidth]{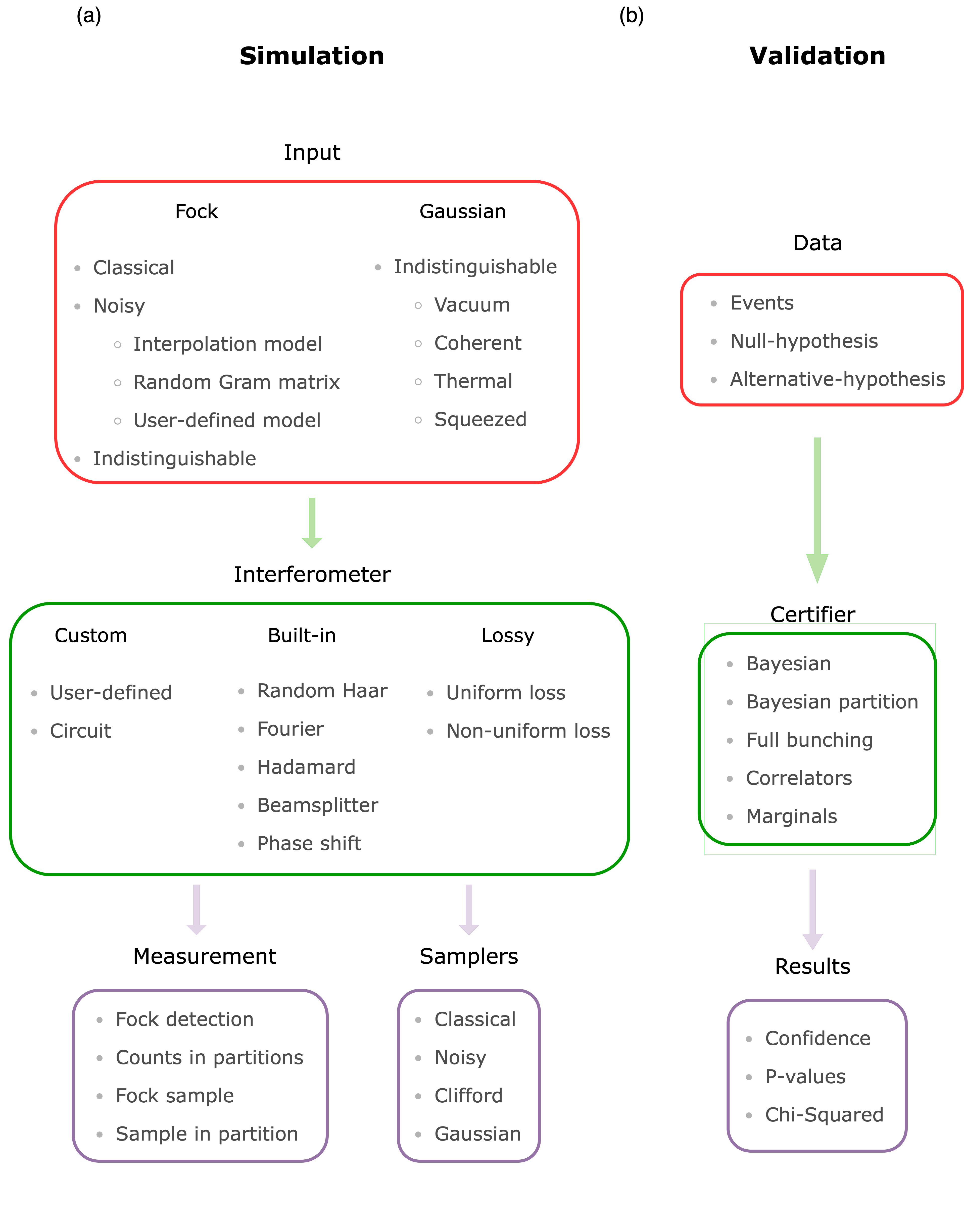}
    \caption{Architecture of \textsc{BosonSampling.jl} (a) Simulation framework: the abstract types representing the experimental setup are displayed as the colored boxes, containing they related concrete types. (b) Validation framework. An extensive description of every type and their usage can be found in the \href{https://benoitseron.github.io/BosonSampling.jl/stable/}{documentation} and the associated tutorials.}
    \label{fig:archi}
\end{figure}

The present package takes advantage of Julia's type system through a hierarchy used to classify the building blocks of a quantum interferometry experiment, namely: a photonic \texttt{Input} state sent through an \texttt{Interferometer} and measured according to an \texttt{OutputMeasurement}. This last category also includes (classical) sampling.
Those three elements are represented by abstract types, that is, they cannot be instantiated and form the backbone of the package structure. 
They can be seen as containers of sets of related concrete types, which can be instantiated. 
Concrete types will represent the specific building blocks of a given interferometric setup. 
For instance, an interferometer defined from a Haar distributed unitary, \texttt{RandHaar}, is a concrete type of the abstract type \texttt{Interferometer} as displayed in Fig. \ref{fig:archi}.

The latter structure is at the core of the code efficiency and tunability. Indeed, the generic programming allowed by Julia combined to the inherent type hierarchy makes the package easy to adapt depending on the user's needs. 
As a new model can simply be embedded as a concrete type of an existing abstract type, implementing new elements can be done easily without modifying the overall package structure, or any function that acts in the same conceptual fashion on the new type. For instance, a function adding noise could work similarly on any type of \texttt{Interferometer}, be it \texttt{RandHaar} or \texttt{Hadamard}.
We will now describe the three main abtract building blocks used in simulating an experiment.


\paragraph{Input}
When considering photons in Fock states at the input, \textsc{BosonSampling.jl} offers three families of input types depending on the partial distinguishability of the particles. We refer to indistinguishable photons (identical arrival time, polarization) using \texttt{Bosonic}. When they are partially distinguishable, they fall under \texttt{PartDist}. Finally, we use \texttt{Distinguishable} for completely distinguishable input photons. 
The type \texttt{PartDist} allows for a general description of the partial distinguishability. Common models used in the literature are also implemented as subtypes \cite{tichy2015_partial_distinguishability}.

The input state can also be \texttt{Gaussian}. In order to take into account partial distinguishability, we adopt the model introduced in \cite{https://doi.org/10.48550/arxiv.2105.09583}. More precisely, any Gaussian input will be model of fully indistinguishable states while, partial distinguishability can be introduced at the level of the interferometer. The input state is decomposed into interfering and non-interfering modes which evolve independently through the interferometer.

\paragraph{Interferometers}
The second building block of a quantum multi-photon experiment is the interferometer. A common practice when simulating boson sampling is sample a unitary matrix according to the Haar random measure. Specific interferometers, such as the discrete Fourier and the Hadamard transforms are built-in, together with elementary linear optical circuit elements such as beamsplitters or phase shifts. 
This allows the user to build networks from scratch.
Generic interferometers are supported.
Loss can be accounted for in generality. 


\paragraph{Measurement/Sample}
This object represents the output of the experiment, which we classify between a measurement that can be the probability of a specific output pattern, property or a randomly generated sample.
For Fock states, a possibility is to observe a given output mode occupation $\vect{s} = (s_1, s_2,...)$, where $0 \leq s_i \leq n$ is the number of photons in mode $i$. Likewise, we can observe a specific partition photon count $\vect{k} = (k_1, k_2,...)$, where $0 \leq k_i \leq n$ is the number photons in a bin $K_i$ gathering multiple output modes.

Different samplers are available depending on the level of distinguishability of the input particles or the use of a lossy interferometer. 
For exact sampling, we provide Clifford-Clifford's algorithm \cite{cliffords}. For imperfect sampling, we rely on the works of Moylett et al. \cite{Moylett_2019}.
Their detailed implementations are discussed in the package documentation.

\subsubsection{Validation}

Experimental data can be analysed within this package to validate boson sampling devices. The procedure is straightforward for the user: data is held in form of \texttt{Events}, comprising all known experimental parameters, such as the detector outputs. Hypotheses are provided, such as that the input is made of indistinguishable particles. An alternative hypothesis could be that they are distinguishable. Another task that can be addressed with this package is to estimate the amount of noise, e.g due to partial distinguishability.

Common validation protocols are implemented, with a comprehensive statistical analysis, providing p-values, etc. A detailed description of these protocols can be found in \cite{walschaers2020signatures, validation, shchesnovich2019noise, flamini2020validating}.


\section{Examples}
This section presents some basic experiments that can be simulated by \textsc{BosonSampling.jl}. Further details about the package, as well as a complete, step-by-step \href{https://benoitseron.github.io/BosonSampling.jl/stable/tutorial/installation.html}{tutorial}, can be found in the \href{https://benoitseron.github.io/BosonSampling.jl/stable/}{documentation}, hosted on the \href{https://github.com/benoitseron/BosonSampling.jl}{GitHub} page of the package. As for any registered Julia package, \textsc{BosonSampling.jl} can be installed by simply executing the following commands in the Julia REPL:
\begin{code}
\begin{Shaded}
\begin{Highlighting}[]
\ImportTok{using} \BuiltInTok{Pkg}
\BuiltInTok{Pkg}\NormalTok{.}\FunctionTok{add}\NormalTok{(}\StringTok{"BosonSampling"}\NormalTok{)}
\end{Highlighting}
\end{Shaded}

\clabel{code:Installation}
\ccaption{code:Installation}{\emph{Installing \textsc{BosonSampling.jl}}.}
\end{code}

\subsection{Hong-Ou-Mandel}
It is well known since the experimental work of Hong, Ou and Mandel \cite{HOM} that two indistinguishable photons impinging on the two input modes of a balanced beamsplitter will bunch in one of the two output modes, leaving a zero probability to observe one photon at each output mode (the coincidence probability). As an introductory example, we review the effect of partial distinguishability in the HOM effect. In this context, it is a common practice to use the time delay $\Delta \tau$ between the two incoming beams as a source of partial distinguishability, while the distinguishability parameter itself is $\Delta\omega\Delta\tau$ with $\Delta\omega$ the uncertainty of the frequency distribution. In order to make the parallel with the built-in interpolating model, we substitute its linear parameter $x$ by $e^{-\Delta\omega^2\Delta\tau^2}$. In this way, a distinguishable input is recovered for $\Delta\omega\Delta\tau \rightarrow \infty$ and a bosonic input for $\Delta\tau=0$. 

\begin{figure}
\centering
\begin{subfigure}{.25\textwidth}
    \includegraphics[width=\textwidth]{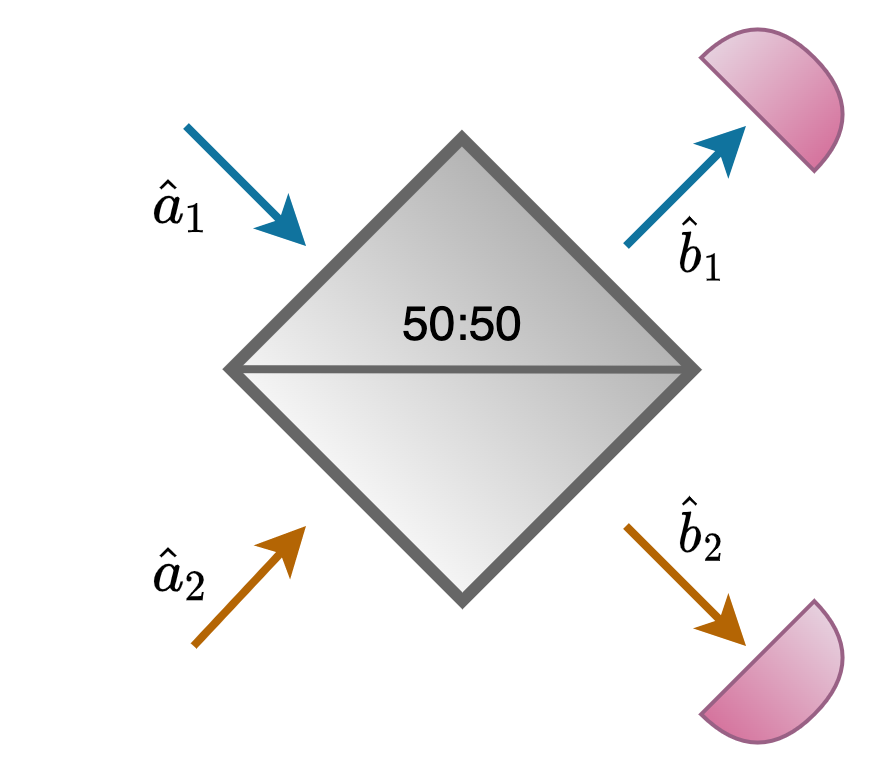}
    \caption{}
    \label{fig:HOM_setup}
\end{subfigure}
\begin{subfigure}{.22\textwidth}
    \includegraphics[width=\textwidth]{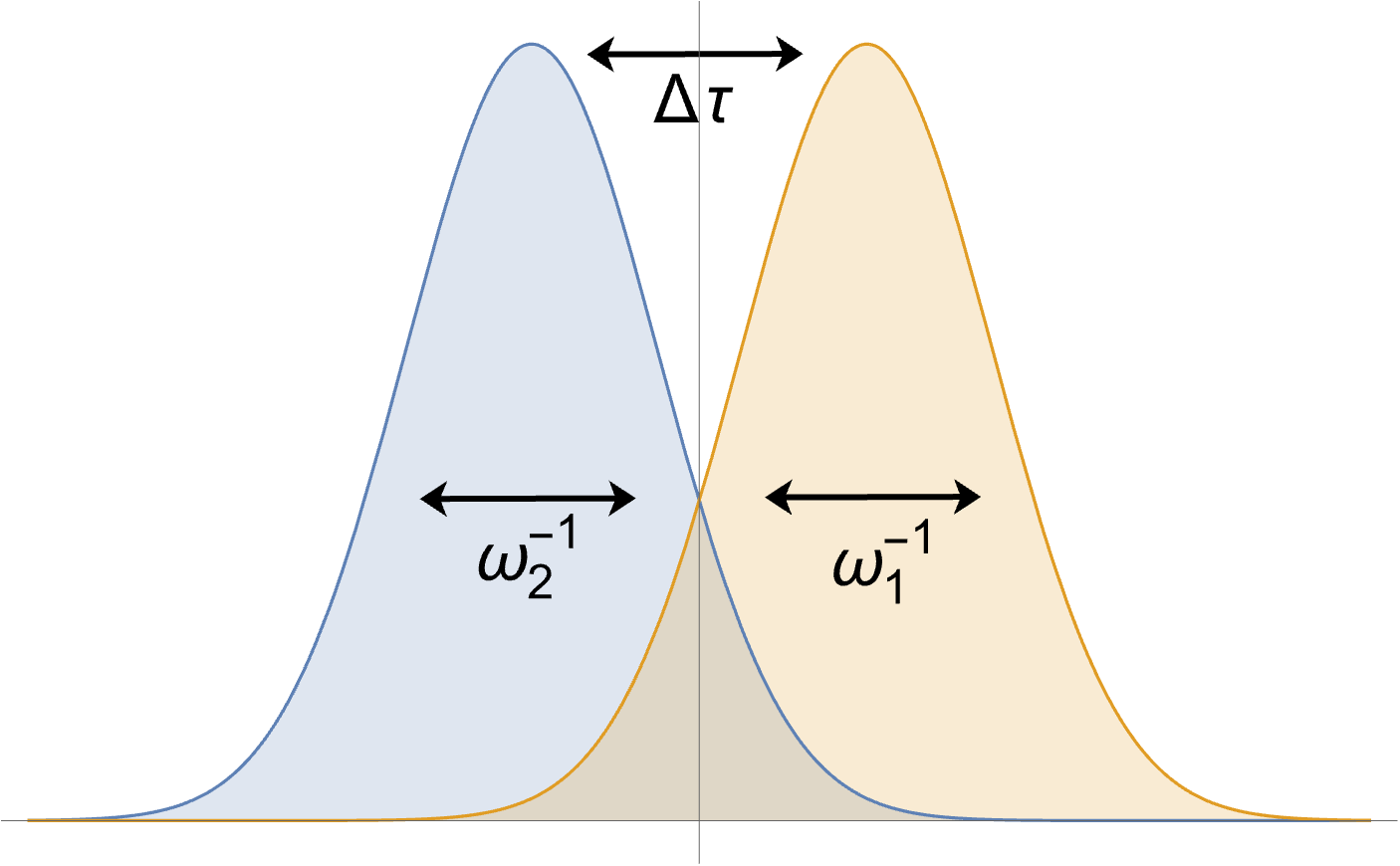}
    \caption{}
    \label{fig:HOM_overlap}
\end{subfigure}
\begin{subfigure}{.35\textwidth}
    \includegraphics[width=\textwidth]{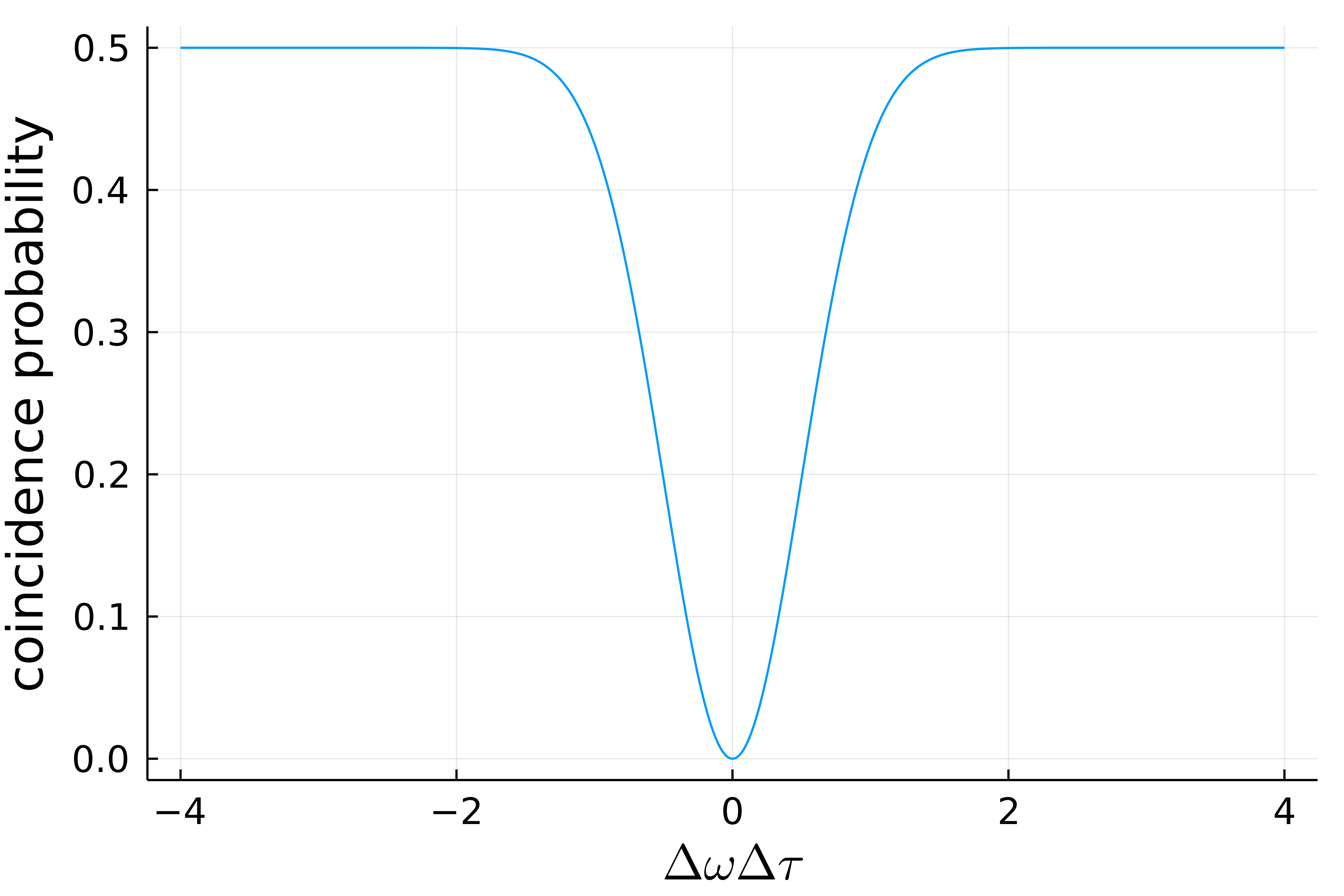}
    \caption{}
    \label{fig:HOM_dip}
\end{subfigure}
        
\caption{The Hong-Ou-Mandel effect: (a) experimental setup (b) time delay between the incoming wave packets (c) HOM dip translating the bosonic interference appears when the overlap between the beams is perfect.}
\label{fig:HOM}
\end{figure}

We reproduce in the \coderef{code:HOM-effect} the HOM experiment as described in Fig. \ref{fig:HOM_setup}, with a variable overlap between the two initial wave functions (Fig. \ref{fig:HOM_overlap}) and compute the coincidence probability. 
\begin{code}
\begin{Shaded}
\begin{Highlighting}[]
\CommentTok{\# Set experimental parameters}
\NormalTok{delta\_omega }\OperatorTok{=} \FloatTok{1}
\CommentTok{\# Set the model of partial distinguishability}
\NormalTok{T }\OperatorTok{=}\NormalTok{ OneParameterInterpolation}
\CommentTok{\# Define the balanced beams{-}plitter}
\NormalTok{B }\OperatorTok{=} \FunctionTok{BeamSplitter}\NormalTok{(}\FloatTok{1}\OperatorTok{/}\FunctionTok{sqrt}\NormalTok{(}\FloatTok{2}\NormalTok{))}
\CommentTok{\# Set each particle in a different mode}
\NormalTok{r\_i }\OperatorTok{=} \FunctionTok{ModeOccupation}\NormalTok{([}\FloatTok{1}\NormalTok{,}\FloatTok{1}\NormalTok{])}

\CommentTok{\# Define the output as detecting a coincidence}
\NormalTok{r\_f }\OperatorTok{=} \FunctionTok{ModeOccupation}\NormalTok{([}\FloatTok{1}\NormalTok{,}\FloatTok{1}\NormalTok{])}
\NormalTok{o }\OperatorTok{=} \FunctionTok{FockDetection}\NormalTok{(r\_f)}

\CommentTok{\# Will store the events probability}
\NormalTok{events }\OperatorTok{=}\NormalTok{ []}

\ControlFlowTok{for}\NormalTok{ delta\_t }\KeywordTok{in} \OperatorTok{{-}}\FloatTok{4}\OperatorTok{:}\FloatTok{0.01}\OperatorTok{:}\FloatTok{4}
    \CommentTok{\# distinguishability}
\NormalTok{    dist }\OperatorTok{=} \FunctionTok{exp}\NormalTok{(}\FunctionTok{{-}}\NormalTok{(delta\_omega }\OperatorTok{*}\NormalTok{ delta\_t)}\OperatorTok{\^{}}\FloatTok{2}\NormalTok{)}
\NormalTok{    i }\OperatorTok{=} \FunctionTok{Input}\DataTypeTok{\{T\}}\NormalTok{(r\_i,dist)}

    \CommentTok{\# Create the event}
\NormalTok{    ev }\OperatorTok{=} \FunctionTok{Event}\NormalTok{(i,o,B)}
    \CommentTok{\# Compute its probability to occur}
    \FunctionTok{compute\_probability!}\NormalTok{(ev)}

    \CommentTok{\# Store the event and its probability}
    \FunctionTok{push!}\NormalTok{(events, ev)}
\ControlFlowTok{end}
\end{Highlighting}
\end{Shaded}

\clabel{code:HOM-effect}
\ccaption{code:HOM-effect}{\emph{Effect of Partial distinguishability in the Hong-Ou-Mandel effect}.}
\end{code}



\subsection{Sampling}
As described in Sec. \ref{overview}, several classical algorithms for boson sampling are implemented, such as the Clifford and Clifford's algorithm \cite{cliffords} for exact sampling and algorithms taking into account typical sources of noise when considering realistic schemes \cite{Moylett_2019}. 
To simulate a boson sampling experiment, one defines an input, the interferometer and a measurement. 
As in the previous example, an \texttt{Event} type is used as a container to hold all interferometric parameters.

\begin{code}
\begin{Shaded}
\begin{Highlighting}[]
\NormalTok{n }\OperatorTok{=} \FloatTok{20}
\NormalTok{m }\OperatorTok{=} \FloatTok{400}

\CommentTok{\# Define an input of 20 photons among 400 modes}
\NormalTok{i }\OperatorTok{=} \FunctionTok{Input}\DataTypeTok{\{Bosonic\}}\NormalTok{(}\FunctionTok{first\_modes}\NormalTok{(n,m))}

\CommentTok{\# Define the interferometer}
\NormalTok{interf }\OperatorTok{=} \FunctionTok{RandHaar}\NormalTok{(m)}

\CommentTok{\# Set the output measurement}
\NormalTok{o }\OperatorTok{=} \FunctionTok{FockSample}\NormalTok{()}

\CommentTok{\# Create the event}
\NormalTok{ev }\OperatorTok{=} \FunctionTok{Event}\NormalTok{(i, o, interf)}

\CommentTok{\# Simulate}
\FunctionTok{sample!}\NormalTok{(ev)}
\CommentTok{\# output:}
\CommentTok{\# state = [0,1,0,...]}
\end{Highlighting}
\end{Shaded}

\clabel{code:sampler}
\ccaption{code:sampler}{\emph{Simulate a perfect boson sampling experiment involving 20 single-photon Fock states in 400 modes via Clifford and Clifford's algorithm}.}
\end{code}

\begin{figure}[h]
    \centering
    \includegraphics[width=0.48\textwidth]{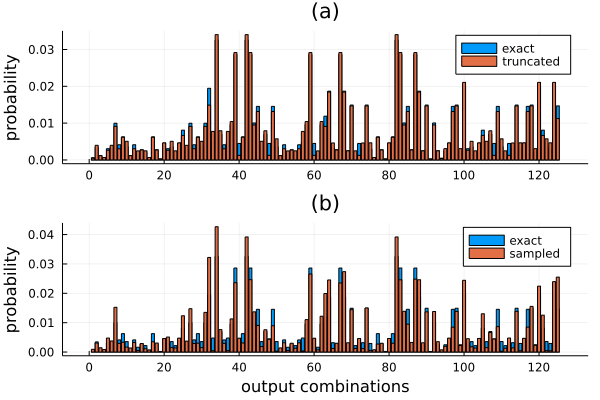}
    \caption{Photon-counting statistics at the output of a 5-mode Haar distributed unitary matrix. (a) Comparison between the ideal distribution and the approximated one. (b) Comparison between the theoretical distribution and a sampled distribution from $10^5$ samples.}
    \label{full_stat}
\end{figure}

Given $n$ photons and $m$ modes, the package proposes methods to compute the entire photon-counting distribution exactly
or approximately when different sources of noise are included \cite{renema2018classical} through the \texttt{noisy\textunderscore distribution} function. 
It returns by default the exact distribution, an approximated distribution in which the computation of the probabilities are truncated and a distribution reconstructed from a Metropolis Independent Sampler (MIS). Below we compute those three distributions for $n=3$ photons in $m=5$ modes with a distinguishability parameter $x=0.74$ and a loss $l=0.63$ (see Fig. \ref{full_stat}).

\begin{code}
\begin{Shaded}
\begin{Highlighting}[]
\CommentTok{\# Set the model of partial distinguishability}
\NormalTok{T }\OperatorTok{=}\NormalTok{ OneParameterInterpolation}

\CommentTok{\# Define an input of 3 photons placed}
\CommentTok{\# among 5 modes}
\NormalTok{x }\OperatorTok{=} \FloatTok{0.74}
\NormalTok{i }\OperatorTok{=} \FunctionTok{Input}\DataTypeTok{\{T\}}\NormalTok{(}\FunctionTok{first\_modes}\NormalTok{(}\FloatTok{3}\NormalTok{,}\FloatTok{5}\NormalTok{), x)}

\CommentTok{\# Interferometer}
\NormalTok{l }\OperatorTok{=} \FloatTok{0.63}
\NormalTok{U }\OperatorTok{=} \FunctionTok{RandHaar}\NormalTok{(i.m)}

\CommentTok{\# Compute the full output statistics}
\NormalTok{p\_exact, p\_truncated, p\_sampled }\OperatorTok{=}
\FunctionTok{noisy\_distribution}\NormalTok{(input}\OperatorTok{=}\NormalTok{i,loss}\OperatorTok{=}\NormalTok{l,interf}\OperatorTok{=}\NormalTok{U)}
\end{Highlighting}
\end{Shaded}

\clabel{code:full_distr}
\ccaption{code:full_distr}{\emph{Computing the output mode occupation statistics with partial distinguishability and loss for 3 photons at the input of a (5,5) Haar distributed unitary}.}
\end{code}

Both computing \texttt{p\textunderscore truncated} and \texttt{p\textunderscore sampled} have by default a probability of failure and a maximal error equal to $10^{-4}$. It is still possible to refine their computations by setting the keywords \texttt{error} and \texttt{failure\textunderscore probability} when calling \texttt{noisy\textunderscore distribution}.

\subsection{Photon counting in partitions}

One of the novel theoretical tools implemented in this package is the ability to efficiently compute photon counting probabilities in partitions of the output modes of the interferometer. A detailed theoretical and numerical investigation can be found in Ref. \cite{validation}. 

Let us now show how we can obtain the probability of finding $k = 0,...,n$ photons in a subset consisting of the first half of the modes of a Haar-randomly chosen an interferometer with $n = 25$ photons and $m = 400$ modes.
With these parameters, a brute force calculation would be pratically impossible, while the new methods used in this package allow for a fast computation.

\begin{code}
\begin{Shaded}
\begin{Highlighting}[]
\NormalTok{n }\OperatorTok{=} \FloatTok{25}
\NormalTok{m }\OperatorTok{=} \FloatTok{400}

\CommentTok{\# Experiment parameters}
\NormalTok{input\_state }\OperatorTok{=} \FunctionTok{first\_modes}\NormalTok{(n,m)}
\NormalTok{interf }\OperatorTok{=} \FunctionTok{RandHaar}\NormalTok{(m)}
\NormalTok{i }\OperatorTok{=} \FunctionTok{Input}\DataTypeTok{\{Bosonic\}}\NormalTok{(input\_state)}

\CommentTok{\# Subset selection}
\NormalTok{s }\OperatorTok{=} \FunctionTok{Subset}\NormalTok{(}\FunctionTok{first\_modes}\NormalTok{(}\FunctionTok{Int}\NormalTok{(m}\OperatorTok{/}\FloatTok{2}\NormalTok{),m))}
\NormalTok{part }\OperatorTok{=} \FunctionTok{Partition}\NormalTok{(s)}

\CommentTok{\# Want to find all photon counting probabilities}
\NormalTok{o }\OperatorTok{=} \FunctionTok{PartitionCountsAll}\NormalTok{(part)}

\CommentTok{\# Define the event and compute probabilities}
\NormalTok{ev }\OperatorTok{=} \FunctionTok{Event}\NormalTok{(i,o,interf)}
\FunctionTok{compute\_probability!}\NormalTok{(ev)}
\CommentTok{\# About 30s execution time on a single core}
\CommentTok{\#}
\CommentTok{\# output:}
\CommentTok{\#}
\CommentTok{\# 0 in subset = [1, 2,..., 200]}
\CommentTok{\# p = 4.650035467008141e{-}8}
\CommentTok{\# ...}
\end{Highlighting}
\end{Shaded}

\clabel{code:partition_photon_count}
\ccaption{code:partition_photon_count}{This snippet retrieves the probability of finding $k = 0,...,n$ photons in the top half of the output modes, with $n=25$ and $m=400$. Note that this calculation would be completely out of reach using a brute-force approach.}
\end{code}



\subsection{Validation}
\label{sec:validation}

We now show how experimental data can be validated. In this example, we compare the null hypothesis $H_0 = $ (bosonic input) to the alternative $H_a = $ (distinguishable input).
Multiple validation protocols are implemented in this package, as shown for instance in Fig. \ref{fig:archi}. 
We choose to use the new formalism developed by some of the authors of photon counting in binned output modes.


In Fig. \ref{fig:validation}, we display an example of validation protocol. Data is assessed using a Bayesian procedure on the event probabilities (instead of partition photon counting in the code sample). As explained in detail in \cite{flamini2020validating, validation}, each sample updates the confidence level $\xi$. To be satisfied that the null hypothesis is true, we could require $\xi = 95\%$ for instance. We plot this curve averaged over $n_{trials} = 500$ with $n_{samples} = 300$ events for each run, in a randomly chosen interferometer with $m = 30$ modes and $n = 10$ photons. The input photons are set to be indistinguishable.
The heatmap shows the density of the $\xi$-curves, on a logarithmic scale. 
\raggedbottom
\subsection{Incorporating new detectors: dark counts}
\label{darCounts}
We now show one of the main advantages of this package and Julia: the ability to create new models while keeping most of the structure untouched. 

Consider a simple model of threshold detector. With a probability $p$ the detector clicks even though no photon was present, so-called dark count. 

We start by defining a new type of output measurement that we label \texttt{DarkCountFockSample}, which is a sub-type of \texttt{OutputMeasurement}. It takes as arguments the probability $p$ to observe a dark count. 
A variable holds the sampled \texttt{ModeOccupation}, but is first initialised empty at creation of \texttt{DarkCountFockSample}. It will be filled later. This is an interesting feature of Julia: the absence of an argument is integrated efficiently by the compiler while a typical, variable-size array would not.
\begin{code}
\begin{Shaded}
\begin{Highlighting}[]
\KeywordTok{mutable struct}\NormalTok{ DarkCountFockSample}
    \OperatorTok{\textless{}:}\DataTypeTok{ OutputMeasurement}

\NormalTok{    s}\OperatorTok{::}\DataTypeTok{Union\{ModeOccupation, Nothing\}}
\NormalTok{    p}\OperatorTok{::}\DataTypeTok{Real}

    \KeywordTok{function} \FunctionTok{DarkCountFockSample}\NormalTok{(p}\OperatorTok{::}\DataTypeTok{Real}\NormalTok{)}
        \ControlFlowTok{if} \FunctionTok{isa\_probability}\NormalTok{(p)}
            \FunctionTok{new}\NormalTok{(}\ConstantTok{nothing}\NormalTok{, p)}
        \ControlFlowTok{else}
            \FunctionTok{error}\NormalTok{(}\StringTok{"invalid probability"}\NormalTok{)}
        \ControlFlowTok{end}
    \KeywordTok{end}
\KeywordTok{end}
\end{Highlighting}
\end{Shaded}

\clabel{code:dark_counts1}
\ccaption{code:dark_counts1}{Implementation of dark counts in the detectors in the \textsc{BosonSampling} framework.}
\end{code}

The second step is to define a new method \texttt{sample!} that implements a sampling algorithm for this specific type of detector. 

As mentioned before, Julia enables multiple dispatch, which means that the new method will not overwrite the already existing ones but will extend the method to the new detectors.
That is, when calling \texttt{sample!}, Julia will automatically apply the relevant method for a given type (\texttt{FockSample}, \texttt{DarkCountFockSample},...).
The new algorithm first extracts a sample without dark counts, then adds their influence on the photon counts: 
\begin{code}
\begin{Shaded}
\begin{Highlighting}[]
\KeywordTok{function} \FunctionTok{sample!}\NormalTok{(ev}\OperatorTok{::}\DataTypeTok{Event\{TIn,TOut\}}\NormalTok{)}
    \KeywordTok{where}\NormalTok{ \{TIn }\OperatorTok{\textless{}:}\DataTypeTok{ InputType}\NormalTok{,}
\NormalTok{           TOut }\OperatorTok{\textless{}:}\DataTypeTok{ DarkCountFockSample}\NormalTok{\}}

    \CommentTok{\# sample without dark counts}
\NormalTok{    ev\_no\_dark }\OperatorTok{=} \FunctionTok{Event}\NormalTok{(ev.input\_state,}
                       \FunctionTok{FockSample}\NormalTok{(),}
\NormalTok{                       ev.interferometer)}

    \FunctionTok{sample!}\NormalTok{(ev\_no\_dark)}
\NormalTok{    sample\_no\_dark }\OperatorTok{=}
\NormalTok{    ev\_no\_dark.output\_measurement.s}

    \CommentTok{\# add count with probablity p}
    \FunctionTok{observe\_dark\_count}\NormalTok{(p) }\OperatorTok{=}
    \FunctionTok{Int}\NormalTok{(}\FunctionTok{do\_with\_probability}\NormalTok{(p))}
\NormalTok{    dark\_counts }\OperatorTok{=}
\NormalTok{    [}\FunctionTok{observe\_dark\_count}\NormalTok{(ev.output\_measurement.p)}
\NormalTok{    for i }\KeywordTok{in} \FloatTok{1}\OperatorTok{:}\NormalTok{ ev.input\_state.m]}

\NormalTok{    ev.output\_measurement.s }\OperatorTok{=}
\NormalTok{    sample\_no\_dark }\OperatorTok{+}\NormalTok{ dark\_counts}
\KeywordTok{end}
\end{Highlighting}
\end{Shaded}

\clabel{code:dark_count2}
\ccaption{code:dark_count2}{Defining a new sampling algorithm.}
\end{code}

Now, our new measurement can be used like any other in the previous examples 

\begin{code}
\begin{Shaded}
\begin{Highlighting}[]
\NormalTok{n }\OperatorTok{=} \FloatTok{10}
\NormalTok{m }\OperatorTok{=} \FloatTok{10}
\NormalTok{p\_dark }\OperatorTok{=} \FloatTok{0.1}
\NormalTok{input\_state }\OperatorTok{=} \FunctionTok{first\_modes}\NormalTok{(n,m)}
\NormalTok{interf }\OperatorTok{=} \FunctionTok{RandHaar}\NormalTok{(m)}
\NormalTok{i }\OperatorTok{=} \FunctionTok{Input}\DataTypeTok{\{Bosonic\}}\NormalTok{(input\_state)}
\NormalTok{o }\OperatorTok{=} \FunctionTok{DarkCountFockSample}\NormalTok{(p\_dark)}
\NormalTok{ev }\OperatorTok{=} \FunctionTok{Event}\NormalTok{(i,o,interf)}

\FunctionTok{sample!}\NormalTok{(ev)}
\CommentTok{\# output:}
\CommentTok{\# state = [3, 1, 0, 3, 0, 1, 2, 0, 0, 1]}
\end{Highlighting}
\end{Shaded}

\clabel{code:dark_count3}
\ccaption{code:dark_count3}{Usage of the newly added detectors.}
\end{code}

\section{Benchmarks}

\subsection{Julia is fast!}
\begin{figure}[h]
    \centering
    \includegraphics[width=0.43\textwidth]{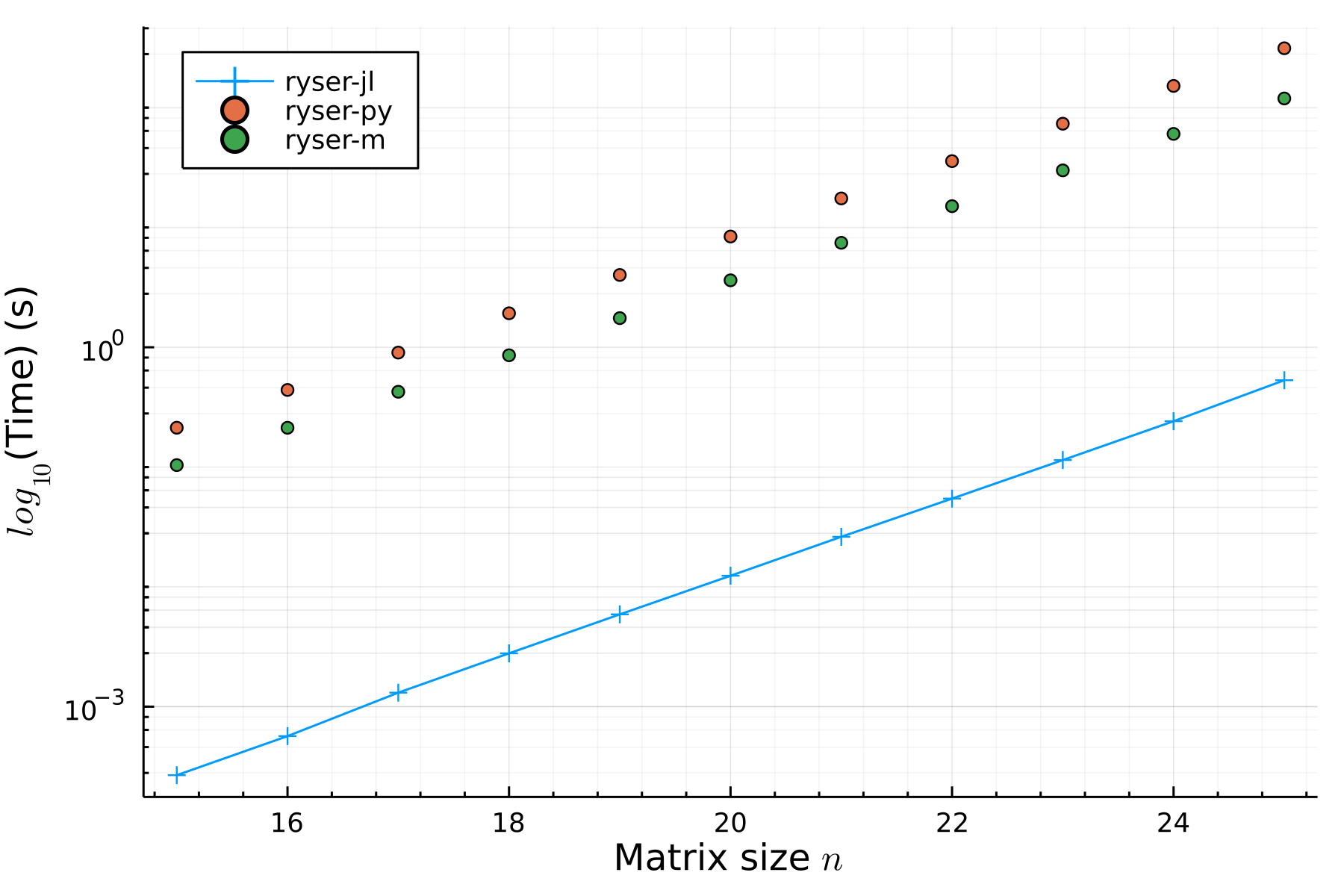}
    \caption{Benchmarks for the single-core running time of Ryser's algorithm in Julia, Matlab and Python on a 3.2GHz Apple M1 processor.}
    \label{fig:ryser_runtime}
\end{figure}

When simulating a boson sampling experiment, the most time consuming part comes from the computation of the transition probabilities, since they are related to the evaluation of the permanent. The best known exact algorithm for its computation is due to Ryser \cite{leech_1964} and runs in $\mathcal{O}\left(2^{n-1}n\right)$ time for a matrix of size $n$ (using Gray ordering). Having an intensive usage of Ryser's algorithm, we compare the running time of its implementation in Julia with two other interpreted languages:  Matlab and Python. Fig. \ref{fig:ryser_runtime} shows how Julia outperforms the two other implementations by two orders of magnitude in computation time.
 
\subsection{Comparison to existing softwares}
The present package is designed for multi-photon interference and boson sampling experiments, taking into account sources of noise and providing a wide variety of validation tools while remaining highly flexible to new paradigms. Other packages for the numerical evaluations of matrix functions and simulation bosonic systems were published in the last few years. Among the most popular softwares are \textsc{The Walrus} \cite{Gupt2019} and  \textsc{Perceval} \cite{https://doi.org/10.48550/arxiv.2204.00602}, both written in Python with a C++ backend.

\begin{figure}[H]
    \centering
    \includegraphics[width=0.67\textwidth]{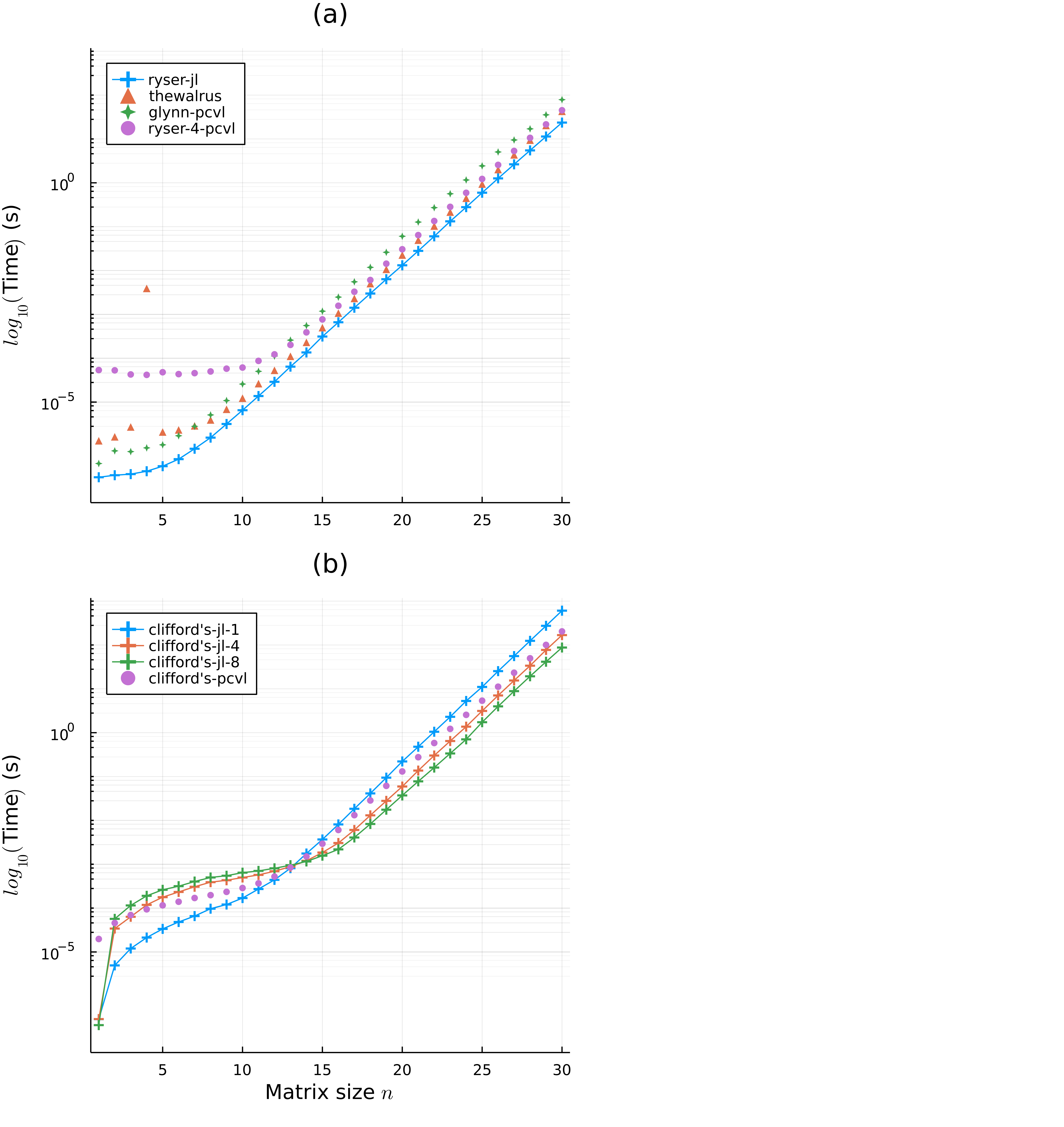}
    \caption{(a) Benchmarks measuring the average time elapsed when computing the permanent of matrices with increasing size $n$. Are displayed the running times of (orange curve) \textsc{The Walrus}, version 0.19, on a single-core, (blue curve) Ryser implementation of \textsc{Permanent.jl}, version 0.1.1, single-core, (green dots) Glynn implementation of \textsc{Perceval}, version 0.6.2 (Quandelibc version 0.5.3), on a single-core, (purple dots) Ryser implementation of \textsc{Perceval}, 4 threads. (b) Average time elapsed when simulating perfectly indistinguishable bosons via Clifford and Clifford's algorithm in Julia on 1, 4 and 8 threads compared to \textsc{Perceval}'s implementation.
    Both benchmarks are realized with the same configuration as in Fig. \ref{fig:ryser_runtime}.}
    \label{fig:my_label}
\end{figure}

We illustrate how the Julia programming language is efficient and how, indeed, it can reach C-like performance while keeping the coding as easy, fast and convenient as in Python. 
We benchmark common functionalities in the three softwares. We first compute permanents of Haar distributed unitary matrices\footnote{Benchmarks inspired from \url{https://the-walrus.readthedocs.io/en/latest/gallery/permanent_tutorial.html} and \url{https://github.com/Quandela/Perceval/blob/main/scripts/performance.py}} (Fig. \ref{fig:my_label}.a) showing that our implementation of Ryser's algorithm remains faster even on single-core. Our simulations of perfect boson sampling through Clifford and Clifford's algorithm achieves comparable performance to what is possible on \textsc{Perceval}, and is more performant only via multithreading (Fig. ~\ref{fig:my_label}.b). 

In the final stages of writing this paper, \textsc{SOQCS}~\cite{https://doi.org/10.48550/arxiv.2208.03250} was released, a C++ library centered around performance at cost of handiness and ease of modification. 

\section{Development path}
This package is a continuously expanding, and many more features are expected to be implemented in the coming years. Current development areas are Gaussian boson sampling related functionalities, in particular simulating GBS with partial distinguishability in a lossy channel.  
Many of the functions used in this package will be made faster, such as computing permanents and hafnians via GPU computing, and HPC-friendly deployment through multithreading and distributed computing.

Potential users are welcome to contact the authors for more details or specific development paths, either through the emails listed above or directly through the Github interface.

Outside contributions are welcome, and multiple improvement paths are outlined on the Github discussion page.

\section{Conclusion}

In this paper we introduced a new, free open source Julia package for the high-performance simulation of multi-photon interference.
We motivated how the choice of language is particularly relevant for scientists, and allows for a package structure that is easy to modify while keeping execution time as fast as in low-level languages. 
This package is therefore aimed at experimentalists, who need to implement specifics of their hardware and at theoreticians, who want to develop new models and boson sampling paradigms.
We showed how the performance positively compares to similar packages written in other languages.

\section*{Acknowledgments}

The authors would like to thank Fulvio Flamini for critical reading of the manuscript. We thank Christoph Hotter and David Plankensteiner for discussions and providing the \LaTeX{} template to display Julia code. 
Leonardo Banchi is also thanked for discussions and his contribution to the Permanents.jl package, a basis for the package presented in this article. 
Leonardo Novo and Nicolas Cerf are thanked for discussions.

B.S. is a Research Fellow of the Fonds National de la Recherche Scientifique – FNRS.  A.R is part of the AppQInfo MSCA ITN which received funding from the EU Horizon 2020 research and innovation program under the Marie Sklodowska-Curie grant agreement No 956071.

\bibliographystyle{unsrtnat}
\bibliography{references}

\end{document}